\documentclass[aps,prl,showpacs,twocolumn]{revtex4}
\usepackage{amssymb}
\usepackage{amsmath}
\usepackage{graphicx}

\begin{document}

\title{Interplay between electron band-anticrossing and charge-density-wave
instabilities}
\date{\today}

\begin{abstract}

Our measurements of the Hall coefficient in rare-earth tritelluride
compounds reveal a strong hysteresis between cooling and warming in the low
temperature range where a second unidirectional charge density wave (CDW)
occurs. We show that this effect results from the interplay between two
instabilities: band crossing of the Te $p_{x}$ and $p_{y}$ orbitals at the
Fermi level and CDW, which have a close energy gain and compete. Calculation
of the electron susceptibility at the CDW wave vector with and without band
anticrossing reconstruction of the electron spectrum yields a satisfactory
estimation of the temperature range of the hysteresis in Hall effect
measurements.
\end{abstract}

\author{P.D.~Grigoriev$^{1,2,3}$, A.A.~Sinchenko$^{4,5}$, P.A.~Vorobyev$^{4}$%
, A.~Hadj-Azzem$^{6}$, P.~Lejay$^{6}$,  A.~Bosak$^7$ and P.~Monceau$^{6}$}

\address{$^{1}$L. D. Landau Institute for Theoretical Physics, 142432,
	Chernogolovka, Russia}

\address{$^{2}$National University of Science and Technology "MISiS", 119049
Moscow, Russia}

\address{$^{3}$P.N.	Lebedev Physical Institute, RAS, 119991, Moscow, Russia}

\address{$^{4}$M.V. Lomonosov Moscow State University, 119991, Moscow,
	Russia}

\address{$^{5}$Kotelnikov Institute of Radioengineering and Electronics of
	RAS, 125009, Moscow, Russia}

\address{$^{6}$Univ. Grenoble Alpes, Inst. Neel, F-38042 Grenoble, France,
	CNRS, Inst. Neel, F-38042 Grenoble, France}

\address{$^{7}$ESRF - The European Synchrotron, 71, Avenue des Martyrs, F-38000 Grenoble, France}

\maketitle

Crossing of electron energy bands near the Fermi level, resulting in the
degeneracy and anticrossing of energy levels, always leads to amazing
physical properties. The anticrossing of spin-split energy bands with
spin-orbit coupling produces non-trivial topologically-protected electron
states in Weyl and Dirac semimetals, which is a subject of extensive
research last decade \cite%
{HasanRMP2010,ZhangRMP2011,YanAnnuRev2017,ArmitageRMP2018}. Even without
spin effects, the band anticrossing near the Fermi level modifies the
electron spectrum and the Fermi surface (FS). This affects various
electronic instabilities such as, superconductivity in high-temperature
cuprate superconductors \cite{LSCO17,LSCO18,LSCO98} and spin- or
charge-density waves \cite{Ru08,Brouet08,EuTe4}. In this paper we unveil the
competition of band anticrossing and the charge-density wave (CDW) in the
family of rare-earth tritelluride compounds. We show, both theoretically and
experimentally, that this interplay leads to the hysteretic electronic phase
transition with the change of FS topology and of Hall coefficient.

Layered compounds of $R$Te$_{3}$ family ($R$=rare earth atom) have a weakly
orthorhombic crystal structure (space group $Cmcm$). These systems exhibit
an incommensurate CDW through the whole $R$ series \cite%
{Ru08,Lavagnini10R,Moore2010}, with a wave vector $\mathbf{Q}%
_{CDW1}=(0,0,\sim 2/7c^{\ast})$ and a Peierls transition temperature above
300 K for the light atoms (La, Ce, Nd). For the heavier $R$ (Tb, Dy, Ho, Er,
Tm) a second CDW occurs at low temperature with the wave vector $\mathbf{Q}%
_{CDW2}=(\sim 2/7a^{\ast},0,0)$ perpendicular to $\mathbf{Q}_{CDW1}$.

For our study we chose three compounds from the $R$Te$_{3}$ family: two
compounds, ErTe$_{3}$ and HoTe$_{3}$, demonstrating bidirectional CDW
ordering at $T_{CDW1}=270$ and $283$ K and $T_{CDW2}=160$ and $110$ K
correspondingly, and TbTe$_{3}$ revealing an unidirectional CDW at $%
T_{CDW}=336$ K. Single crystals of these compounds were grown by a self-flux
technique under purified argon atmosphere as described previously \cite%
{SinchPRB12}. Thin samples with a typical thickness 1-3 $\mu $m having a
rectangular shape were prepared by micromechanical exfoliation of relatively
thick crystals glued on a sapphire substrate.

The magnetic field was applied parallel to the $b$ axis. The Hall resistance 
$R_{xy}(B)=[V_{xy}(+B)-V_{xy}(-B)]/I$ was recorded using the van der Pauw method \cite{Pauw61}%
, sweeping the field between $+6$ and $-6$ T at fixed temperature with a
step $\Delta T=10$ K first by cooling from $T>T_{CDW1}$ down to 4.2 K and
after that by warming back.

\begin{figure*}[tbh]
\includegraphics[width=0.3\textwidth]{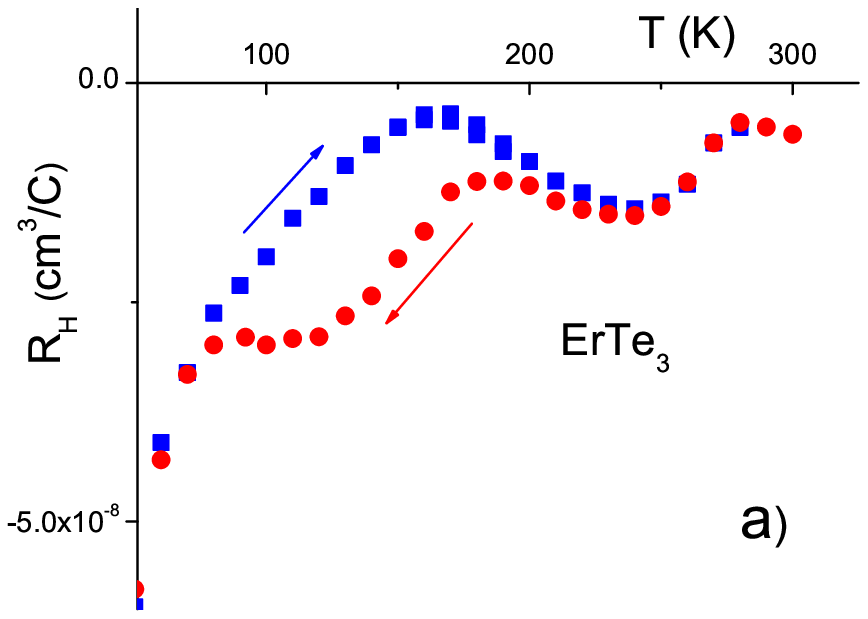} \ \ 
\includegraphics[width=0.3 \textwidth]{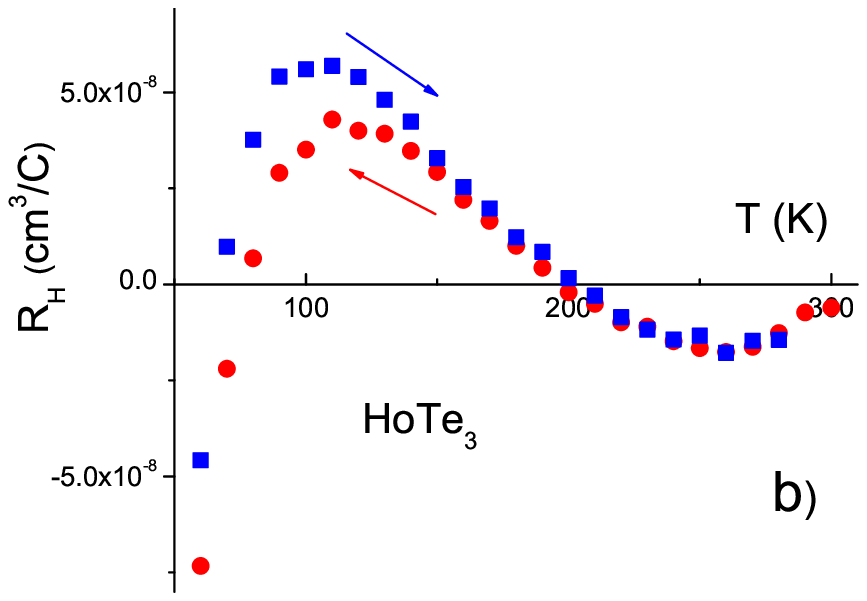} \ \ 
\includegraphics[width=0.3\textwidth]{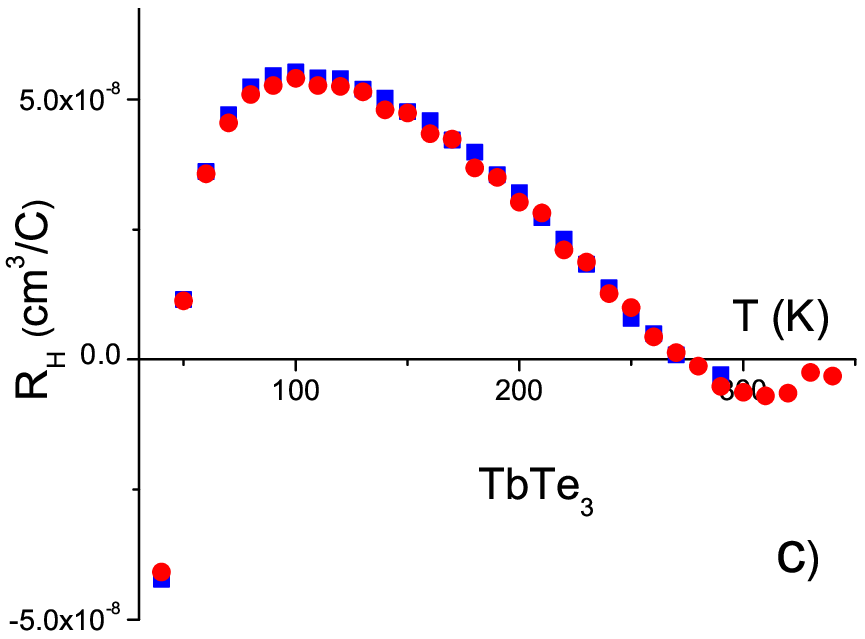}
\caption{(color online) Hall constant of ErTe$_{3}$ (a), HoTe$_{3}$ (b) and
TbTe$_{3}$ (c) as a function of temperature. Red circles correspond to
cooling, and blue squares correspond to warming.}
\label{F1}
\end{figure*}

For all measured compounds $R_{xy}$ is a linear function of $B$ at least for
temperatures $T\gtrsim 100$ K (see Supplementary Materials (SM)). So, the Hall constant, $R_{H}=R_{xy}d/B$,
where $d$ is the crystal thickness, is indeed a field-independent quantity.
Its temperature dependencies for ErTe$_{3}$, HoTe$_{3}$ and TbTe$_{3}$ are
shown in Fig. \ref{F1} (a), (b) and (c) correspondingly. One can see that
for ErTe$_{3}$ and HoTe$_{3}$ $R_{H}$ demonstrates a strong hysteresis
between cooling and warming in the temperature range around the second
Peierls transition while $R_{H}(T)$ is completely reversible for TbTe$_{3}$
revealing only a single transition to the CDW state in the studied range of
temperature. When measured under cooling and warming the temperature
dependence of the resistance $R(T)$ of all three compounds was reversible (see SM).
It means that the total number of charge carriers remains near the same
under cooling and warming. We see only one explanation of this effect: there
are two types of carriers, and the hysteresis observed is attributed to the
change in electron-hole balance as a result of the second CDW formation.
Such scenario is confirmed by the change of the sign of Hall constant at a
certain temperature in HoTe$_{3}$ and TbTe$_{3}$.

One can naturally attribute the observed effect to the hysteresis of the CDW
wave vector $\boldsymbol{Q}_{CDW}$ due to its pinning by crystal
imperfections. However, our preliminary x-ray diffraction studies of ErTe$%
_{3}$, performed at ID28 ESRF beamline \cite{Girard19}, showed a completely
reversible evolution of all structural parameters in the temperature range
100-300 K (see SM). Therefore, we consider another possible origin
of this hysteresis, based on the interplay of CDW$_{2} $ with another type
of electronic instability. As a possible candidate of such electronic
ordering competing with CDW$_2$, we suggest the one due to the electron
band-crossing at the Fermi level.

\begin{figure}[tbh]
\includegraphics[width=7cm]{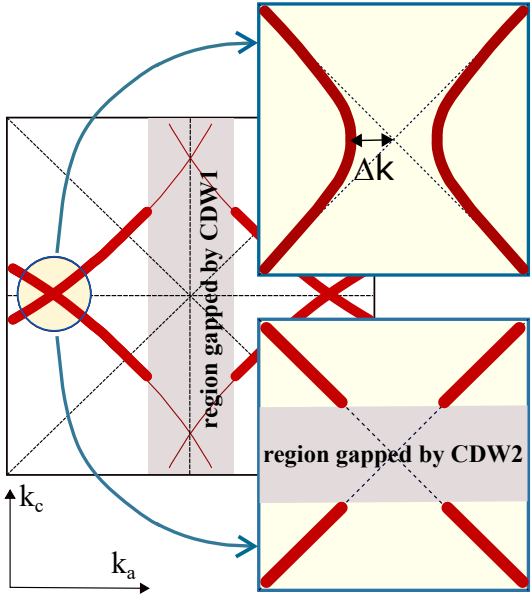}
\caption{\textbf{Schematic representation of the Fermi surface (FS) in RTe$%
_{3}$.} Main figure: the full FS with the CDW$_{1}$-gapped region shaded by
gray. In the crossing region, highlighted by a filled yellow circle, two FS
reconstructions are possible: with band anticrossing shown in the right
upper inset, and with energy gap due to CDW$_2$ shaded by gray in the lower
right inset.}
\label{FS0}
\end{figure}

Consider two electron bands with electron dispersion $\epsilon _{1}\left( 
\boldsymbol{k}\right) $ and $\epsilon _{2}\left( \boldsymbol{k}\right) $.
Two corresponding Fermi surfaces, given by equations $\epsilon _{1}\left( 
\boldsymbol{k}\right) =E_{F}$ and $\epsilon _{2}\left( \boldsymbol{k}\right)
=E_{F}$, intersect along the lines $\{\boldsymbol{k}_{0}\}$ in the momentum
space. In RTe$_{3}$ compounds in the $\left( k_{x},k_{y}\right) $ plane
below $T_{CDW1}$ there are two such crossing points $\boldsymbol{k}_{0}$,%
\cite{CommentCP} highlighted by filled yellow circles in Fig. \ref{FS0}. At
each degeneracy point $\boldsymbol{k}_{0}$, any small interband coupling $%
V\left( \boldsymbol{Q}\right) $, even at zero momentum transfer $\boldsymbol{%
Q}=0$, leads to the band anticrossing and to the reconstruction of FS (see
right upper inset in Fig. \ref{FS0}). This FS anticrossing has been observed
in various RTe$_{3}$ compounds by ARPES measurements \cite%
{Moore2010,ARPES2008}. The interband coupling $V\left( \boldsymbol{Q}\right) 
$ may originate, e.g., from the electron-electron (e-e) interaction.
Usually, $\left\vert V\left( \boldsymbol{Q}\right) \right\vert $ decreases
with the increase of momentum transfer $\left\vert \boldsymbol{Q}\right\vert 
$, and $\left\vert V\left( \boldsymbol{0}\right) \right\vert \equiv V_{0}$
may considerably exceed $\left\vert V\left( \boldsymbol{Q\neq 0}\right)
\right\vert $.

\begin{figure*}[tbh]
\includegraphics[width=0.3\textwidth]{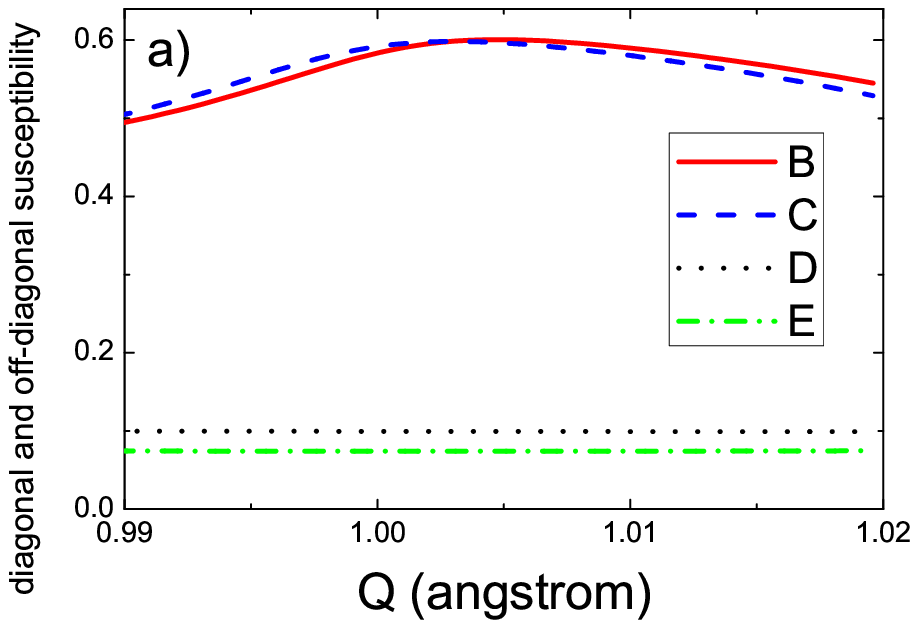} \ \ 
\includegraphics[width=0.3\textwidth]{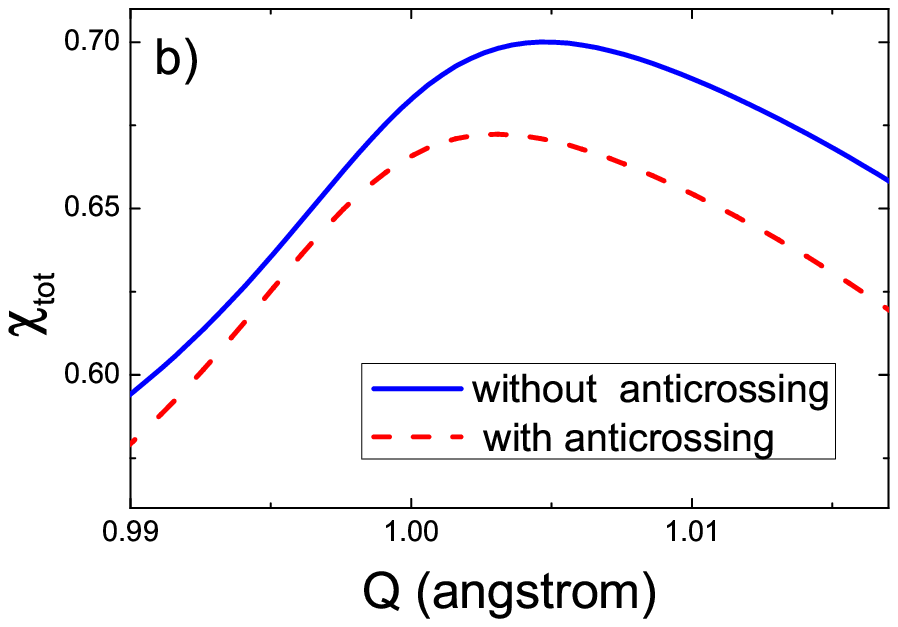} \ \ 
\includegraphics[width=0.3\textwidth]{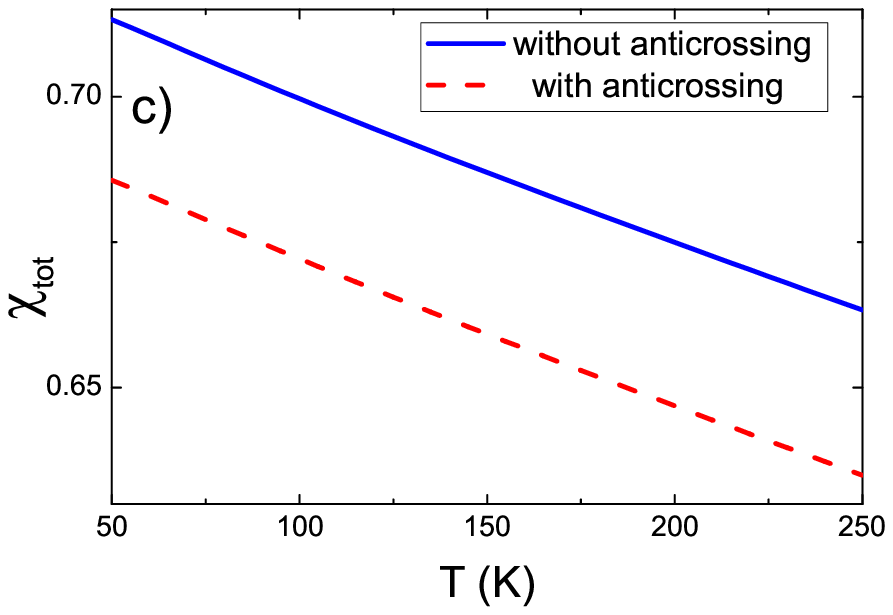}
\caption{\textbf{Calculated electron susceptibility.} (a) The electron
susceptibility contributions without (solid blue and green lines) and with
the band-crossing reconstruction (dashed red and black lines). The upper
two curves (blue and red) show the calculated "diagonal" intraband
contributions $\protect\chi $, while two lower curves (green and black)
show "off-diagonal" interband contributions $\Delta \protect\chi $ to
susceptibility. (b) Total susceptibility $\protect\chi _{total}=\protect\chi %
+\Delta \protect\chi $ as a function of wave vector $Q_{x}$ near its maximum
without (solid blue line) and with the band-crossing reconstruction (dashed
red line). (c) The temperature dependence of maximal total susceptibility
without (solid blue line) and with the band-crossing reconstruction (dashed
red line).}
\label{FigS}
\end{figure*}

First consider the toy model with the interband coupling (off-diagonal
terms) only at $\boldsymbol{Q}=0$. In this model the different momenta are
not coupled, and the Hamiltonian writes down as a sum over electron momenta $%
\boldsymbol{k}$, $\hat{H}=\sum_{\boldsymbol{k}}\hat{H}_{\boldsymbol{k}}$,
where in the basis of two branches $\alpha =1,2$ of electron spectrum each
term $\hat{H}_{\boldsymbol{k}}$ is given by a $2\times 2$ matrix\cite{NoteH}%
\begin{equation}
\hat{H}_{\boldsymbol{k}}=\left( 
\begin{array}{cc}
\epsilon _{1}\left( \boldsymbol{k}\right) & V_{0} \\ 
V_{0} & \epsilon _{2}\left( \boldsymbol{k}\right)%
\end{array}%
\right)  \label{Hk}
\end{equation}%
with two eigenvalues 
\begin{equation}
E_{\pm }\left( \boldsymbol{k}\right) =\frac{\epsilon _{1}\left( \boldsymbol{k%
}\right) +\epsilon _{2}\left( \boldsymbol{k}\right) }{2}\pm \sqrt{\left( 
\frac{\epsilon _{1}\left( \boldsymbol{k}\right) -\epsilon _{2}\left( 
\boldsymbol{k}\right) }{2}\right) ^{2}+V_{0}^{2}},  \label{Epm}
\end{equation}%
representing two new branches of electron spectrum. The total electron
energy is given by the sum of quasiparticle energies over their quantum
numbers: 
\begin{equation}
\mathcal{E}=\sum_{\boldsymbol{k},\alpha }E_{\alpha }\left( \boldsymbol{k}%
\right) n_{F}\left[ E_{\alpha }\left( \boldsymbol{k}\right) \right] ,
\label{E1}
\end{equation}%
where $n_{F}(\varepsilon )=1/\left( 1+\exp \left[ (\varepsilon -E_{F})/T%
\right] \right) $ is the Fermi-Dirac distribution function. Without band
anticrossing the total energy $\mathcal{E}_{0}$ is given by the same Eq. (%
\ref{E1}) with the replacement $E_{\alpha }\left( \boldsymbol{k}\right)
\rightarrow \epsilon _{\alpha }\left( \boldsymbol{k}\right) $. The
difference $\Delta \mathcal{E}_{AC}=\mathcal{E}-\mathcal{E}_{0}$ comes
mainly from the vicinity of the crossing points $\boldsymbol{k}_{0}$, where
two conditions are satisfied: (i) $\left\vert \epsilon _{1}\left( 
\boldsymbol{k}\right) -\epsilon _{2}\left( \boldsymbol{k}\right) \right\vert
\lesssim V_{0}$, so that the electron spectrum changes considerably, and
(ii) $\left\vert \epsilon _{1}\left( \boldsymbol{k}\right) +\epsilon
_{2}\left( \boldsymbol{k}\right) \right\vert \lesssim 2V_{0}$, so that the
change of electron spectrum is close to the Fermi level. Near the crossing
point $\boldsymbol{k}_{0}$ one may linearize each branch of electron
spectrum: 
\begin{equation}
\epsilon _{\alpha }\left( \boldsymbol{k}\right) \approx \boldsymbol{v}%
_{F\alpha }\left( \boldsymbol{k}-\boldsymbol{k}_{0}\right) ,  \label{epsLin}
\end{equation}%
where $\boldsymbol{v}_{F\alpha }$ is the Fermi velocity $v_{F}$ of branch $%
\alpha $. Then $\left\vert \epsilon _{1}\left( \boldsymbol{k}\right) \pm
\epsilon _{2}\left( \boldsymbol{k}\right) \right\vert \approx \left( 
\boldsymbol{v}_{F1}\pm \boldsymbol{v}_{F1}\right) \left( \boldsymbol{k}-%
\boldsymbol{k}_{0}\right) \sim v_{F}\left\vert \boldsymbol{k}-\boldsymbol{k}%
_{0}\right\vert $, and the contributing momentum area in the vicinity of the
crossing point $\boldsymbol{k}_{0}$ is estimated as $\left(
V_{0}/v_{F}\right) ^{2}$. Then the energy difference per unit area per one
spin component but including two cross points is 
\begin{equation}
\Delta \mathcal{E}_{AC}\sim -\left( V_{0}/\pi \hbar v_{F}\right)
^{2}V_{0}\approx -V_{0}^{3}\,a^{2}\rho _{F}^{2},  \label{DE1}
\end{equation}%
where $a$ is the in-plane lattice constant and $\rho _{F}=1/\pi \hbar v_{F}a$
is the quasi-1D density of states (DoS) at the Fermi level per one branch
and spin component. Our calculation of $\Delta \mathcal{E}_{AC}$ by the
numerical integration according to Eq. (\ref{E1}) confirms the estimate in
Eq. (\ref{DE1}), giving the $\Delta \mathcal{E}_{AC}$ value $20$\% less than
in Eq. (\ref{DE1}).

The CDW energy gain is\cite{Gruner,CommentE}%
\begin{equation}
\Delta \mathcal{E}_{CDW}=-\Delta ^{2}\rho _{F},  \label{E2}
\end{equation}%
where $\Delta $ is the CDW energy gap, given by the off-diagonal matrix
element of the Hamiltonian, similar to $V_{0}$ in Eq. (\ref{Hk}). The extra
small parameter $\eta \equiv V_{0}\rho _{F}a^{2}=V_{0}a/\pi \hbar v_{F}\ll 1$
in the band-crossing energy gain in Eq. (\ref{DE1}) as compared to Eq. (\ref%
{E2}) comes from the small momentum region of contributing electrons, while
in a CDW a considerable part of electrons on the Fermi level participate in
the Peierls instability, so that a similar small factor $a^{2}\rho
_{F}\Delta $ does not appear. Hence, the CDW$_{2}$ energy gain may be larger
than the energy gain from band anticrossing, although its energy gap $\Delta
_{2}\ll V_{0}$.

We estimate the value of $V_{0}$ from the FS distortion at the crossing
point $\boldsymbol{k}_{0}$\ observed in ARPES. This FS distortion $\Delta k$
along the $x$-axis is about $3$\% of the Brillouin zone width $2\pi \hbar /a$%
,\cite{Moore2010,ARPES2008} where the lattice constant $a=4.28\mathring{A}$
in ErTe$_{3}$. This $\Delta k$ corresponds to the condition $\left\vert
\epsilon _{1}\left( \boldsymbol{k}\right) +\epsilon _{2}\left( \boldsymbol{k}%
\right) \right\vert =2V_{0}$, giving the boundary of electron states with a
gap on the Fermi level according to Eq. (\ref{Epm}). In RTe$_{3}$ compounds
the FS of two bands cross at almost right angle, as shown in Fig. \ref{FS0}.
Substituting the electron dispersion (\ref{epsLin}) with $v_{F}\approx
1.3\cdot 10^{8}cm/s$, we obtain in ErTe$_{3}$ compounds $V_{0}\approx
v_{F}\Delta k/\sqrt{2}\approx 250meV$. For comparison, in ErTe$_{3}$ the CDW$%
_{1}$ energy gap $\Delta _{1}\approx 175meV$, and the CDW$_{2}$ energy gap
is $\Delta _{2}\approx 55meV$.\cite{CDWGap} The parameter $\eta =V_{0}a/\pi
\hbar v_{F}\approx 0.04$ is indeed $\ll 1$, and the ratio of energy gains
from the band anticrossing and from CDW$_{2}$ is $\Delta \mathcal{E}%
_{AC}/\Delta \mathcal{E}_{CDW2}\approx \eta V_{0}^{2}/\Delta _{2}^{2}\approx
0.8$, i.e. slightly less than unity. This means a strong
temperature-dependent interplay of these two electronic instabilities,
making CDW$_{2}$ slightly more energetically favorable at low $T$. However,
since $V_{0}/\Delta _{2}\approx 5\gg 1$, the band anticrossing appears at
much higher temperature than $T_{CDW2}$, even higher than $T_{CDW1}$.

The band anticrossing and CDW$_{2}$ hinder each other, because each of them
change the electron spectrum. The CDW$_{2}$ creates an energy gap on the
Fermi level just at the spots of FS intersection (see lower right inset in
Fig. \ref{FS0}), thus suppressing or making irrelevant the band
anticrossing. The influence of band anticrossing on CDW$_{2}$ is less
obvious, because the FS has an approximate nesting property both with and
without the band anticrossing. Moreover, our calculation of the DoS with and
without band anticrossing gives nearly the same result in both cases. Hence,
to substantiate that band anticrossing hinders the CDW$_{2}$ instability, we
need to compare the electronic susceptibility $\chi (\boldsymbol{Q},T)$ at
the CDW$_{2}$ wave vector $\boldsymbol{Q}$ in both cases: with and without
band anticrossing reconstruction of electron spectrum. The CDW$_{2}$
transition temperature $T_{c}$ is given by the equation\cite{Gruner} $%
\left\vert U\chi (\boldsymbol{Q}_{max},T_{c})\right\vert =1$, where $%
\boldsymbol{Q}_{max}$ is the wave vector where the susceptibility $\chi $
takes a maximum value. The larger is the susceptibility $\chi $, the higher
is the CDW transtition temperature, because susceptibility increases with
the decrease of temperature.

For calculation we use the well-known formula for the static susceptibility
of free-electron gas at finite wave vector $\boldsymbol{Q}$. Electron spin
only leads to a factor $4$ in susceptibility, but the summation over band
index $\alpha $ must be retained. Then the real part of electron
susceptibility is 
\begin{equation}
\chi \left( \boldsymbol{Q}\right) =\sum_{\alpha ,\alpha ^{\prime }}\int 
\frac{4d^{d}\boldsymbol{k}}{\left( 2\pi \right) ^{d}}\frac{n_{F}\left( E_{%
\boldsymbol{k},\alpha }\right) -n_{F}\left( E_{\boldsymbol{k+Q},\alpha
^{\prime }}\right) }{E_{\boldsymbol{k+Q},\alpha ^{\prime }}-E_{\boldsymbol{k}%
,\alpha }},  \label{chi2}
\end{equation}%
where $d$ is the dimension of space. In $R$Te$_{3}$ compounds under study
there are two bands crossing the Fermi level, $\alpha ,\alpha ^{\prime }=1,2$%
, and we may take $d=2$ because the dispersion in the $z$-direction is weak.
Eq. (\ref{chi2}) differs only by the summation over $\alpha $ and $\alpha
^{\prime }$ from the common expression, e.g., given in Eq. (1.7) of Ref. 
\cite{Gruner}.

Taking the tight-binding bare electron dispersion $\epsilon _{1,2}\left( 
\boldsymbol{k}\right) $ commonly used\cite{Brouet08,Sinchenko2014} for $R$Te$%
_{3}$ compounds and given by Eqs. (2) of Ref. \cite{Sinchenko2014}, we
calculate the susceptibility in Eq. (\ref{chi2}) as a function of the wave
vector $\boldsymbol{Q}$\ and temperature $T$ for two cases: without
band-crossing effect, i.e. for bare electron dispersion $\epsilon
_{1,2}\left( \boldsymbol{k}\right) $, and for reconstructed dispersion given
by Eq. (\ref{Epm}). The results are shown in Fig. \ref{FigS}. The
integration over momentum in Eq. (\ref{chi2}) is performed only at $%
k_{x}>k_{x0}\approx 0.29\mathring{A}^{-1}$, because in the momentum region $%
\left\vert k_{x}\right\vert <k_{x0}$ the electron spectrum at the Fermi
level has a large gap $\Delta _{1}$ due to the CDW$_{1}$. The summation over 
$\alpha $ and $\alpha ^{\prime }$ in Eq. (\ref{chi2}) gives four terms: two
intraband terms $\chi $ with $\alpha =\alpha ^{\prime }$ and two interband
terms $\Delta \chi $ with $\alpha \neq \alpha ^{\prime }$. The intraband
"diagonal" terms, enhanced by a rather good FS nesting, are much larger than
the "off-diagonal" interband terms, because the latter correspond to almost
perpendicular FS sheets and do not have such nesting enhancement (see Fig. %
\ref{FigS}a). Hence, the intraband contribution, shown by upper blue and red
curves in Fig. \ref{FigS}a, have a maximum at the CDW$_{2}$ wave
vector $\boldsymbol{Q}$, resulting to a similar maximum on the total
susceptibility in Fig. \ref{FigS}b, while the interband contribution, shown
by lower green and black curves in Fig. \ref{FigS}a, depends weakly on $%
\boldsymbol{Q}$. Nevertheless, the interband contribution is considerable,
being about 20\% of the intraband susceptibility. While the maximum values
of "diagonal" intraband susceptibility terms are weakly affected by the band
anticrossing, the "off-diagonal" interband terms are suppressed by the band
anticrossing reconstruction by more than 20\% (see Fig. \ref{FigS}a). This
can be easily understood by looking at the FS with and without band
anticrossing, shown in Fig. \ref{FS0}. The DoS and the nesting property is
not violated by the band anticrossing, hence, the intraband terms remain
almost the same (only the optimal CDW$_{2}$ wave vector slightly shifts). On
the contrary, after the band anticrossing reconstruction, the FS of different
bands become separated by $\Delta k\sim 3$\% of the Brillouin zone. Two FS
sheets even do not intersect as was without the band anticrossing. Hence,
the interband susceptibility decreases considerably.

We have shown that the band anticrossing and CDW$_{2}$ interfere,
suppressing each other. With temperature decrease the band anticrossing
appears first (at higher temperature) and reduces the CDW$_{2}$ transition
temperature to its observed value $T_{CDW2}$. At lower temperature, when CDW$%
_{2}$ develops and\ the $\Delta _{2}$ increases, since $\left\vert \Delta 
\mathcal{E}_{CDW2}\right\vert >\left\vert \Delta E_{AC}\right\vert $, the
band anticrossing shrinks in favor of CDW$_{2}$. This may happen as a
first-order phase transition, accompanied by a hysteresis. When the
temperature increases again, the CDW$_{2}$ disappears at temperature $%
T_{CDW2}^{\ast }>T_{CDW2}$, because of the changed band-crossing energy
spectrum. This results to a hysteresis seen by the Hall coefficient
sensitive to the FS reconstruction due to $CDW_{2}$. We can estimate how
strong is this hysteresis by looking at the calculated temperature
dependence of susceptibility $\chi \left( T\right) $, shown in Fig. \ref%
{FigS}c. The calculated optimal wave vector $\boldsymbol{Q}$ of the $CDW_{2}$
instability, i.e. of the susceptibility maximum shown in Fig. \ref{FigS}b,
very slightly increases with temperature from $Q_{x}\approx 1.004\mathring{A}%
^{-1}$ at $T=50$K to $Q_{x}\approx 1.006\mathring{A}^{-1}$ at $T=200$K.
Therefore, this change was not observed in X-ray experiment(see SM). 
In Fig. \ref{FigS}c we plot the maximum value of $\chi \left( \boldsymbol{Q%
}\right)$ as a function of temperature $T$ without (solid blue line) and
with (dashed red line) band-crossing reconstruction. They differ by $4.5\%$
only, but since the temperature dependence of susceptibility is also quite
weak, the susceptibility value $\chi _{c}=0.68$, which the red curve reaches
only at $T_{c}\approx 50K$, the blue curve has already at $T_{c}^{\ast
}\approx 175K$. Thus, the expected temperature hysteresis is rather large: $%
\Delta T=T_{c}^{\ast }-T_{c}\approx 125K$.

The proposed interplay between band-crossing and CDW is rather general and
is expected in many other compounds with FS intersection at the nested
parts. For example, similar effect is expected in $R$Te$_4$ family of
compounds, where a large temperature hysteresis of resistance $\Delta T>100K$
has also been observed recently\cite{EuTe4}.

The bilayer splitting of electron spectrum smears the nesting condition.\cite%
{CDWGap} The exact bare electron dispersion $\epsilon _{1,2}\left( 
\boldsymbol{k}\right) $ are unknown. The coupling between two CDWs in the
above analysis is taken into account only by neglecting the contribution
from the states gapped by CDW$_{1}$. These and other factors make the
interplay of CDW$_{2}$ with other instabilities more complicated, but we
expect that the main features of the proposed model remain valid.

To summarize, we observed a strong hysteresis of the Hall coefficient in the
rare-earth triteluride compounds ErTe$_{3}$ and HoTe$_{3}$, having two CDW
phase transitions. We explain this effect by a strong interplay of the low
temperature CDW and the band-anticrossing change of electron spectrum. We
estimate of the temperature range of this hysteresis by calculating the
electron susceptibility at the CDW$_{2}$ wave vector with and without band
anticrossing. The interplay between these two instabilities is proposed and
investigated for the first time and may be relevant to other compounds where
two electron bands cross at the Fermi level.

\acknowledgements

The authors are grateful to the staff of the ID 28 beam line ESRF. 
The work was partially supported by joint grant CNRS and Russian State Fund
for the Basic Research (No. 17-52-150007) and by the Foundation for
Advancement of Theoretical Physics and Mathematics ''BASIS''. P.V. thanks
RFBR grant No. 19-02-01000. P.G. thanks State assignment 0033-2019-0001
''The development of condensed-matter theory''. A.S. thanks State assignment IRE RAS.


\begin{thebibliography}{99}
\bibitem{CommentCP} Above $T_{CDW1}$ there are 4 crossing points, but two of
them disappear after the CDW$_{1}$ energy gap opens.
\end{thebibliography}
\end{document}


\title{Supplementary information on ''Interplay between electron band-crossing and charge-density-wave instabilities''}
\date{\today}

\author{P.D.~Grigoriev$^{1,2,3}$, A.A.~Sinchenko$^{4,5}$, P.A.~Vorobyev$^{4}$, A.~Hadj-Azzem$^{6}$, P.~Lejay$^{6}$, A.~Bosak$^7$ and P.~Monceau$^{6}$}

\address{$^{1}$L. D. Landau Institute for Theoretical Physics, 142432,
	Chernogolovka, Russia}

\address{$^{2}$National University of Science and Technology "MISiS", 119049
Moscow, Russia}

\address{$^{3}$P.N.	Lebedev Physical Institute, RAS, 119991, Moscow, Russia}

\address{$^{4}$M.V. Lomonosov Moscow State University, 119991, Moscow,
	Russia}

\address{$^{5}$Kotelnikov Institute of Radioengineering and Electronics of
	RAS, 125009, Moscow, Russia}

\address{$^{6}$Univ. Grenoble Alpes, Inst. Neel, F-38042 Grenoble, France,
	CNRS, Inst. Neel, F-38042 Grenoble, France}

\address{$^{7}$ESRF - The European Synchrotron, 71, Avenue des Martyrs, F-38000 Grenoble, France}

\maketitle

\section{Hall effect measurements}

\begin{figure}[tbh]
	\includegraphics[width=8cm]{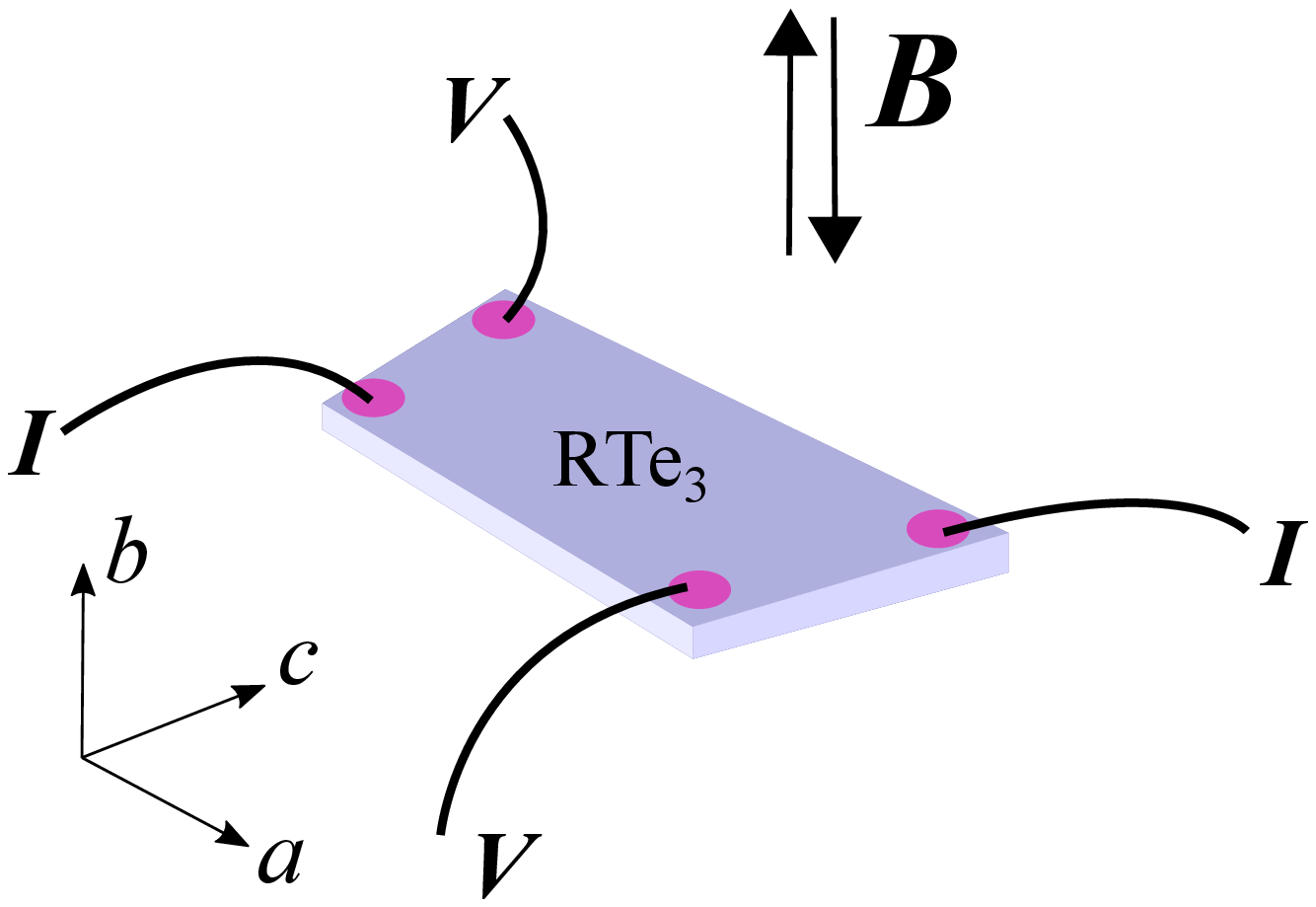}
	\caption{Schematic representation of the Hall effect measurement with the van der Pauw method}
	\label{FS1}
\end{figure}

Schematic representation of the Hall effect measurements is shown in Fig. \ref{FS1}. The magnetic field was applied parallel to the $b$ axis, and the Hall resistance $R_{xy}(B)=[V_{xy}(+B)-V_{xy}(-B)]/2I$ was recorded using the van der Pauw \cite{Pauw61} method, sweeping the field between $+6$ and $-6$ T at fixed temperature with a step $\Delta T=10$ K under the cooling from $T>T_{CDW1}$ up to 4.2 K and after that under warming back.

The shape of a {\it R}Te$_3$ single crystals is a very thin plate with the long b-axis perpendicular to the plane of the
plates. Therefore, for Hall effect measurements the choosen experimental geometry is the most convenient. 

\begin{figure*}[tbh]
	\includegraphics[width=0.3\textwidth]{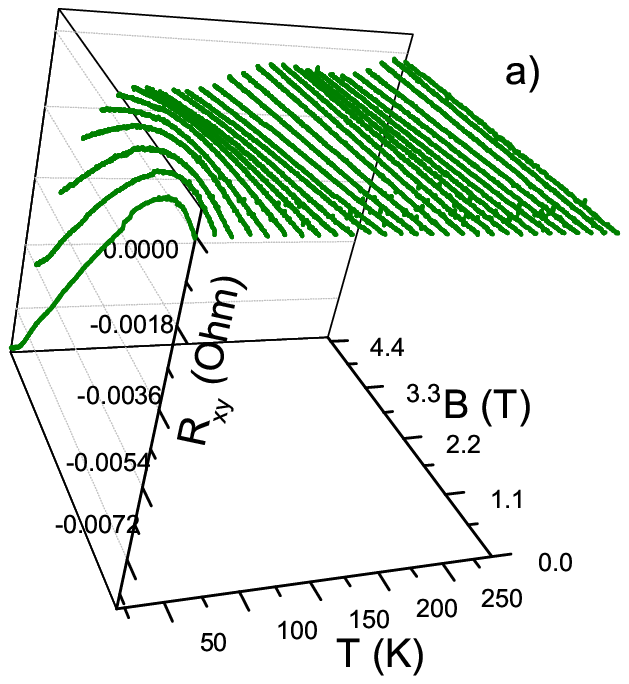} \ \ 
	\includegraphics[width=0.3\textwidth]{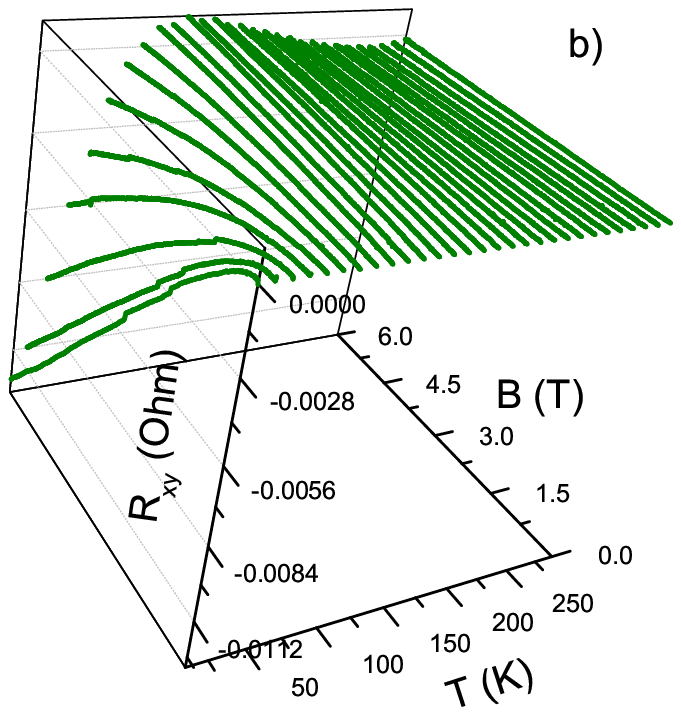} \ \ 
	\includegraphics[width=0.3\textwidth]{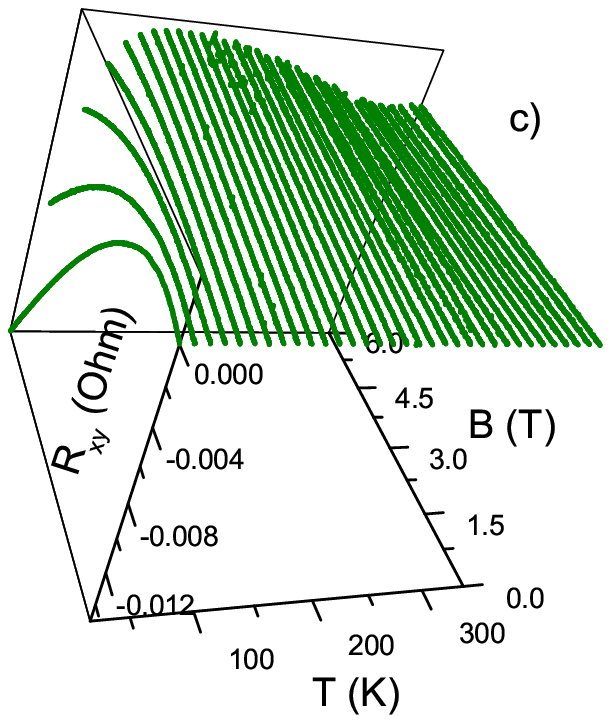}
	\caption{Hall resistance, $R_{xy}$ as a function of temperature and magnetic field of ErTe$_{3}$ (a), HoTe$_{3}$ (b) and
		TbTe$_{3}$ (c) under cooling.}
	\label{FS2}
\end{figure*}

Fig. \ref{FS2} (a-c) represent the Hall resistance as a function of magnetic field in the temperature range from 4.2 K up to the temperature which is well above $T_{CDW1}$ for all studied compounds. Hall resistance changes sign at a certain temperature in HoTe$_{3}$ and TbTe$_{3}$. Such crossing is absent for ErTe$_3$ although it can be seen that $R_H$ also has a tendency to cross zero near 150-170 K (see Fig.1 (a) in main text). However it is the second CDW transition at $T=160$ K which interrupts this tendency. Another fact is that the formal estimation of carrier concentration for the case one band gives unrealistic values, of the order $10^{27}-10^{28}$ cm$^{-3}$. All these facts indicate that there are two types of carriers in {\it R}Te$_3$ compounds with near the same concentration of electrons and holes. 

Usually in such a system with two (electron and hole) conduction channels, the Hall resistivity as a function of $B$ should be non-linear. In this case $\rho_{xy}(B)$ can be well fitted to a two-band model as, 

\begin{equation}
\rho_{xy}(B)=\frac{B}{e}\frac{(n_h\mu_h^2-n_e\mu_e^2)+(n_h-n_e)\mu_e^2\mu_h^2B^2}{e(n_h\mu_h-n_e\mu_e)^2+[(n_h-n_e)\mu_h\mu_eB]^2}
 \label{tbm}
\end{equation}

where $n$ and $\mu$ are respectively carrier density and mobility, and the subscript $e$ (or $h$) denotes electron (or hole).
As can be seen from Fig. \ref{FS2} our $R_{xy}(B)$ dependencies are non-linear in the measured magnetic field range but only at low enough temperatures. At higher temperatures ($>100$ K) $R_{xy}$ is a nearly linear function of $B$ most probably because low enough carriers mobility. Indeed, as can be seen from equation (\ref{tbm}) the non-linearity of $R_{xy}(B)$ is determined mainly by the second term in numerator the value of which strongly depends on carrier mobilities. Estimation from magnetoresistance \cite{Sinch17} gives for $(\mu_e\mu_h)\sim10^3$ cm$^4$/V$^2$s$^2$ that is really small. It can be expected that at higher magnetic field the non-linearity of $R_{xy}(B)$ will be observable at high temperature also. In the case of the present experiment we can consider the Hall constant, $R_H=R_{xy}d$, where $d$ is the crystal thickness, as a real constant value at $T>100$ K.

\section{resistivity}

\begin{figure*}[tbh]
	\includegraphics[width=0.32\textwidth]{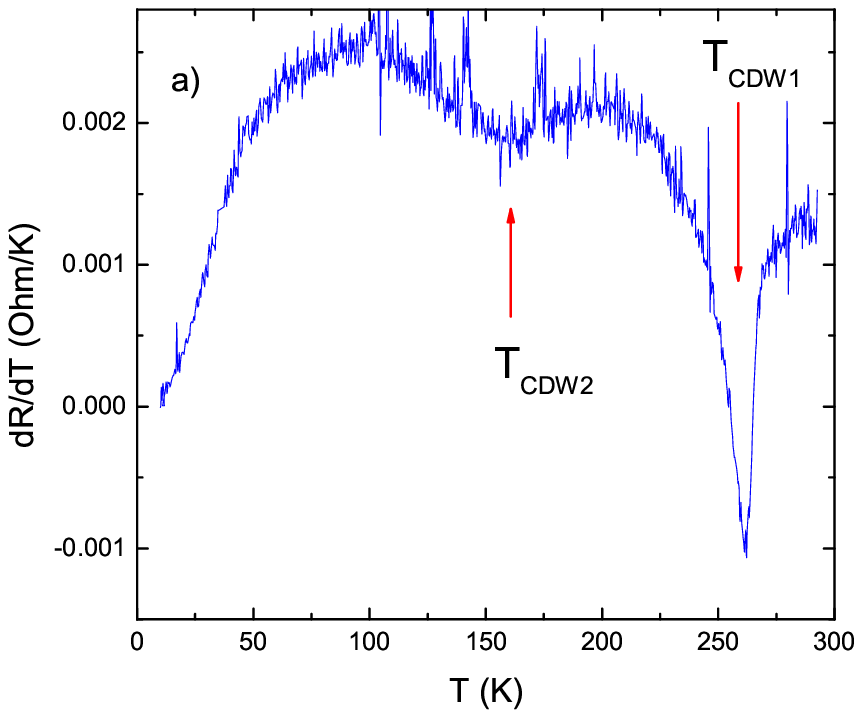} \ \ 
	\includegraphics[width=0.3\textwidth]{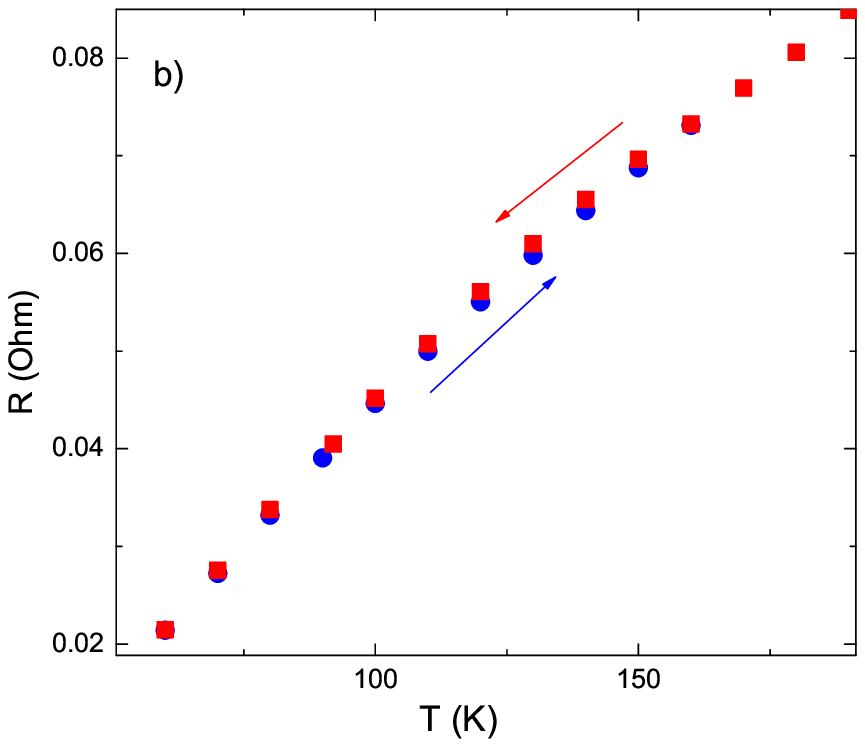} \ \
	\includegraphics[width=0.3\textwidth]{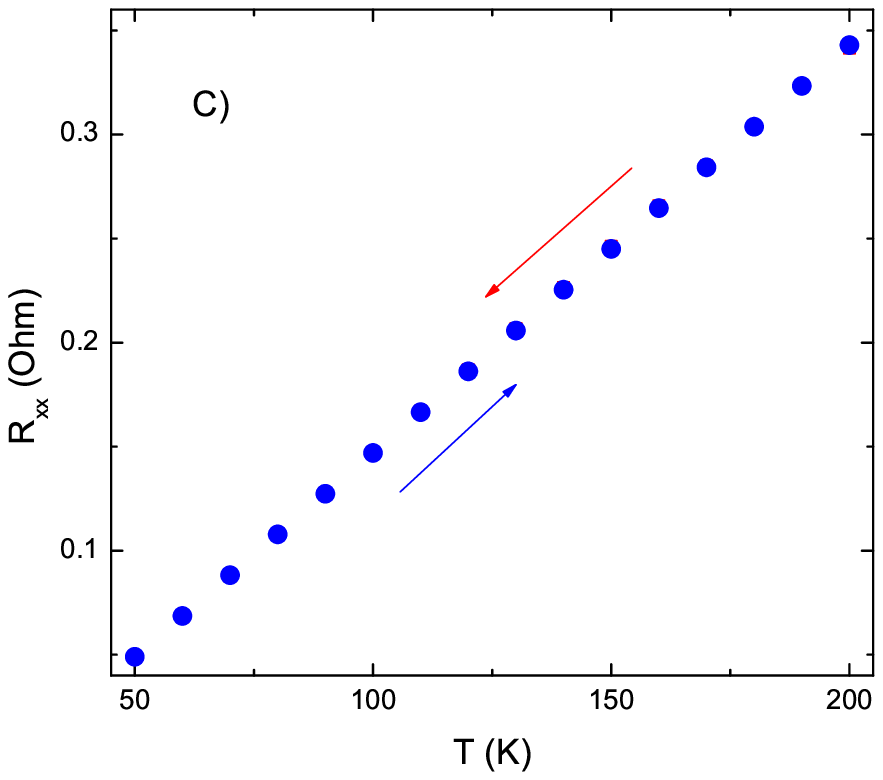}
	\caption{(a) $dR/dT$ as a function of temperature for ErTe$_3$. (b) and (c) temperature dependence of resistance of ErTe$_3$ and HoTe$_3$ under cooling (red) and warming (blue).}
	\label{FS3}
\end{figure*}

The resistivity singularity corresponding to the second CDW transition is very weak in {\it R}Te$_3$ compounds and most often can be resolved only in the derivative $dR/dT(T)$. Fig.\ref{FS3} (a) shows example of such dependency for ErTe$_3$. Fig.\ref{FS3} (b) and (c) demonstrate the behavior of resistance of ErTe$_3$ and HoTe$_3$ samples shown in Fig.1 in main text under cooling and warming in the temperature range
corresponding to the second CDW transition. Each point was taken at fixed $T$ as $R=V(B=0)/I$. As can be seen, this dependence is
reversible in the limit of experimental error. 

\section{X-ray diffraction}

The diffraction measurements were carried out on the side station of the ID28 beamline of the European Synchrotron Radiation Facility (ESRF). We used a monochromatic beam with wavelength of 0.6968 \AA{A}. The frames were acquired with a Pilatus 3X 1M detector in shutterless mode with the angular step of 0.25 deg for several temperatures between room temperature (RT) and 90 K. The sample temperature was controlled by the  Oxford Cryosystems Cryotream 700 Plus. Orientation matrix refinement and preliminary reciprocal space reconstructions were performed using the CrysAlis software package, final reconstructions are produced with locally developed software.

No visible difference was found neither in the intensities of modulation spots nor in the diffuse scattering component for heating and cooling, as illustrated by the reconstructions of h0l, 3kl and hk3 planes shown for 160 K.

\begin{figure}[tbh]
	\includegraphics[width=12cm]{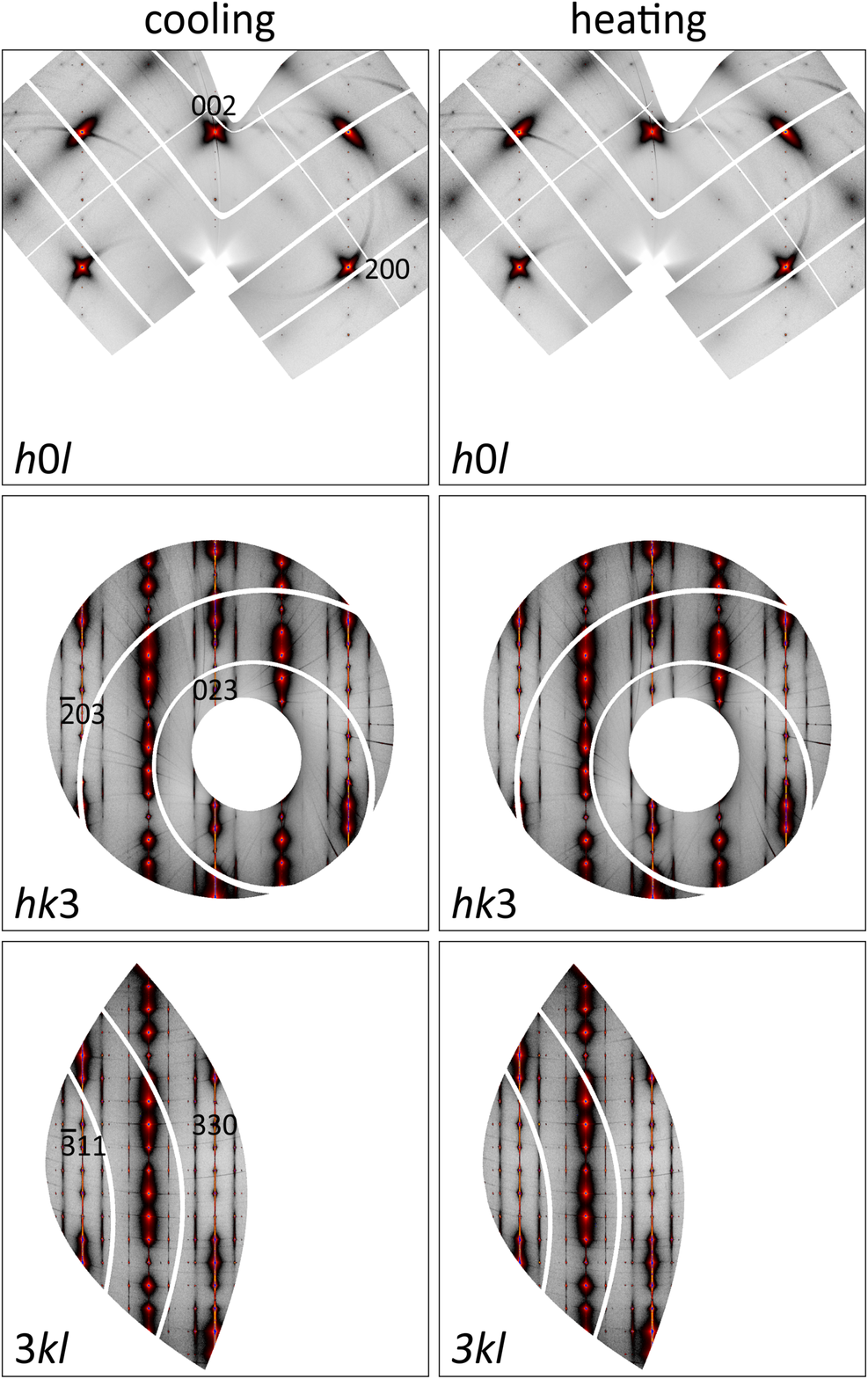} 
		\caption{Reciprocal space planes h0l, 3kl and hk3 of ErTe$_3$ as reconstructed for 160K on cooling from room temperature and as on heating from 90 K.}
	\label{FS4}
\end{figure}